# Genesis of Karl Popper's EPR-Like Experiment and its Resonance amongst the Physics Community in the 1980s

**Flavio Del Santo** –Faculty of Physics, University of Vienna. delsantoflavio@gmail.com

**Abstract**

I present the reconstruction of the involvement of Karl Popper in the community of physicists concerned with foundations of quantum mechanics, in the 1980s. At that time Popper gave active contribution to the research in physics, of which the most significant is a new version of the EPR thought experiment, alleged to test different interpretations of quantum mechanics. The genesis of such an experiment is reconstructed in detail, and an unpublished letter by Popper is reproduced in the present paper to show that he formulated his thought experiment already two years before its first publication in 1982. The debate stimulated by the proposed experiment as well as Popper's role in the physics community throughout 1980s is here analysed in detail by means of personal correspondence and publications.

## 1. Introduction: a fruitful collaboration between physics and philosophy

Quantum Mechanics (QM) has been referred to as "the most successful theory that humanity has ever developed; the brightest jewel in our intellectual crown" (Styer, 2000).

Sir Karl R. Popper (1902-1994) is regarded as "by any measure one of the preeminent philosophers of the twentieth century" (Shield, 2012) and no doubt "one of the greatest philosophers of science" of his time (Thornton 2016).

However, it is not very well known that Popper and QM had an intense and controversial relationship, which lasted for about 60 years. In fact, despite its celebrated success, QM always exhibited a number of problems, which Popper had always grasped.

Since 1934, Popper was indeed one of the foremost critics of the Copenhagen Interpretation of Quantum Mechanics (CIQM), which owes its name to the school of Niels Bohr in Copenhagen, where it was mainly developed. Although providing a proper definition might be controversial, the foundations of CIQM can be summarised in a nutshell as an idealistic, subjectivist and instrumentalist viewpoint on how to interpret quantum formalism. And since its very formulation, this approach was strenuously opposed by some of the most eminent physicists who also had contributed to develop quantum theory: Einstein, Schrödinger and de Broglie among them.

It is worth mentioning that CIQM, acquired an enormous and always increasing consensus throughout the twentieth century, also due to its pragmatic program,[1] which was a straightforward justification for the introduction of a strictly productive exploitation of science in the post-war period.

Nonetheless - after a period of almost complete indifference towards fundamental problems - since the mid-1960s and the formulation of the crucial result of Bell's inequalities (Bell, 1964), the issue of the interpretations of quantum formalism slowly experienced a revival, due to the struggle of a new generation of 'dissidents', who were striving (though with different aims) against the well-established instrumentalist use of physics (see e.g. Freire, 2015; Baracca, Bergia and Del Santo, 2016; Kaiser, 2011).

---

[1] About the pragmatism of CIQM, the Nobel laureate in physics, Sir Anthony Leggett, stressed that it "should probably more correctly be called the Copenhagen non-interpretation" (Leggett, 2012)

For what concerns Popper, he

> fought a lone battle against the Copenhagen interpretation at a time when anyone attempting to criticize orthodoxy was liable to be labelled at best an 'outsider' or at worst a crank. But Popper's carefully argued criticisms won the support of a number of admiring and influential physicists (Redhead, 1985, p. 163)

The present research tries to reconstruct the perhaps overlooked role that Popper played in the scientific debate on Foundations of Quantum Mechanics (FQM) in the 1980s, when he effectively became part of the physics community. In that period Popper published several papers on the subject, also in specialised physics journals, and contributed to international physics conferences. In this respect, they are of central importance the relationships that Popper established (or consolidated) in that decade with some of the most prominent (although dissident) physicists in the field of FQM.

In particular, since 1980 Popper strengthened his friendship with the French physicist Jean-Pierre Vigier (1920-2004), a pupil of Louis de Broglie and a distinguished physicist amongst the critics of orthodox QM.[2] The two of them had a mutual intellectual influence and the remarkable experience arising from their research collaboration throughout the 1980s, which led to two co-authored scientific papers.[3]

Also the physicist Franco Selleri (1936-2013), initiator of the critiques towards QM in Italy,[4] established a close and enduring relation with Popper since 1983, and was certainly influenced by his ideas on QM. Besides the literature of that period, I have analysed their highly influential relationships, thanks to the witnesses of several colleagues of the protagonists in this case study. At the same time, careful research has been conducted to investigate the private correspondence and the personal notes, in Popper's Archive (PA in the references) in Klagenfurt, Austria.[5] Incidentally it ought to be stressed that Selleri's archive in Bari is not yet adequately organised, while most regrettably, Vigier's documents could be forever lost.[6] By means of this documentation, it has been possible to reconstruct in detail (section 3.2) the genesis of the most notable contribution proposed by Popper in physics: "a simplified version of the EPR" thought experiment (Popper, 1982a; the original EPR paper is Einstein, Podolsky and Rosen, 1935), which gave rise to a long-lasting debate within the circles of physicists concerned with FQM. Also, I have retrieved a letter –never published before – containing the first formulation of the mentioned Popper's EPR-like experiment, and dated 1980, that is to say two years prior its first publication.

---

[2] For a short note on the impressive life and career of Vigier, see his obituary written by Holland (2004)

[3] About this, Prof. Joseph Agassi, Popper's pupil and eminent philosopher of science, remarked: "No doubt, Popper's having co-authored a paper with Vigier, is a great honour, since philosophers seldom can cooperate with scientists" (I am grateful to Prof. Agassi for this personal communication on 04/12/2016).

[4] For a detailed reconstruction of Selleri's role in the revival towards the FQM in the post-war Italy see Baracca, Bergia and Del Santo (2016). In that work a short subsection (13.2.2.1) is devoted to a preliminary study of the relations between Selleri and Popper, from which this research has been stimulated.

[5] The *Karl Popper Sammlung* of the Alpen Adria University in Klagenfurt (Austria), collects copies of Popper's correspondences. The original ones are gathered at the Popper Archive in the Hoover Institution at the Stanford University, California. I could retrieve 23 letters from Selleri to Popper, 9 from Popper to Selleri, 6 from Vigier to Popper and 7 from Popper to Vigier, besides many other interesting ones, directed to other famous physicists.

[6] It was with the deepest regret that I learnt from Prof. Cufaro Petroni (letter to the author on 14/11/2016) of the possible irremediable loss of Vigier's documents. In fact, Nicola Cufaro Petroni himself - one of the closest collaborators of Vigier - was commissioned, in February 2004, to sort the documents left in Vigier's office. He could collect three full boxes of (uncatalogued) documents which apparently, due to a renovation, are not traced any more. At the moment, with the kind help of Prof. Cufaro Petroni, we are trying to retrieve the documents which might have a great historical value.

During the 1980s, Popper became a node in a network of physicists, among which the novel ideas (together with the preprints of most influential papers) circulated, were debated and improved to resist the most severe critiques, before in some cases becoming acknowledged science.

Popper had the great merit to have brought together three generations of (quantum) physicists with the philosophy of science, actively participating in the most characteristic periods when criticisms of CIQM were put forward. Foundations of quantum mechanics is the blurred boundary between philosophy and physics, on which Popper has always walked, sometimes attracting physicists inclined towards philosophical interests, yet some other times crossing himself this border into the physicists' domain. In my opinion it is worth reconstructing in detail this peculiar and emblematic case of Popper's involvement into research in physics, when one of those rare contacts between the two worlds - largely considered detached - of physics and philosophy has happened. In fact this could provide a distinctive example of a profitable interdisciplinary collaboration between two subjects that share, to some extent, the same aim, but only seldom interact. Yet, there cannot be good science without carefully considering the interpretations of its models and formulas; on the other hand, philosophy of science obviously requires to be applied to some science to be worthwhile. In this article I have tried to emphasise this particular aspect, reconstructing the period of 1980s, when Karl Popper - a professional, though very influential, philosopher - had the opportunity to actively participate to the debate within the physics community.

**2. Former contacts between Popper and physics**

Although Popper's interest towards QM never diminished throughout his long and distinguished career, it is possible to identify three major periods of his active contribution to (foundations) of quantum physics.

(i) The earliest critiques of Popper towards QM, date back to 1934, when he developed a particular aversion to the CIQM. In particular, a central aspect that Popper could never accept was the Heisenberg's interpretation of the uncertainty relations, which contributes to rule out realism in physics. However it has to be stressed that Popper's objections have never been based on mere philosophical argumentation, nor prejudices, but rather on an empirical and testable basis, accordingly to a methodological position that Popper had been advocating since his early publications (Popper 1934b). Consequently, as early as 1934 he proposed a G*edankenexperiment* aiming at falsifying Heisenberg's relations.[7] The experiment turned out to be "a gross mistake for which I have been deeply sorry and ashamed of ever since", Popper declared later on (Popper, 1982a, p.15). Nevertheless, that thought experiment, though mistaken, gave him the opportunity to actively take part in lively debates directly with Bohr and Heisenberg, initiators of the CIQM, and with Einstein, the most eminent opponent of Copenhagen. It is beyond the aims of the present paper to discuss Popper's contributions to QM in those years, however a short overview can be found in (Shields, 2012) and in Popper's autobiography (Popper, 1976, chapter 18: *Realism and Quantum Theory*). What is notable is that in this first period Popper acquired the friendship and the support of a number of illustrious physicists, among them Alfred Landé, Hermann Bondi and Henry Margenau, and in the 1950s he established contacts with Bohm and had new correspondence and discussions with Einstein and Schrödinger.

(ii) The second entry of Popper in the physics community happened at the end of the 1960s. At that time Popper published a paper (Popper, 1967) with the significant title *Quantum Mechanics without the Observer*, which had a certain resonance among the physicists and gathered many appreciations (e.g. L. De Broglie,

---

[7] This (actually mistaken) thought experiment was firstly proposed in (Popper, 1934a) and also reproduced in Popper's most famous book, *Logik der Forschung* (Popper, 1934b). The first to confute the physical validity of the experiment was of Carl Friedrich von Weizsäcker, followed by Bohr and Einstein themselves. In the English translation *The Logic of Scientific Discovery* (1959, pp. 461-464) is also reproduced a copy of Einstein's critical reply.

Alfred Landé Mario Bunge, Hermann Bondi, Bartel Leendert van der Waerden). It belongs to the same period one of Popper's most debated and influential publications in the highly quoted journal *Nature* (Popper, 1968), in which the author claimed to have found an error in a famous paper by Garrett Birkhoff and John von Neumann. Consequently, Popper entered an intense debate and its conclusions divided the physics community. He especially had to defend his position against the prominent school of logic of quantum mechanics (David Finkelstein, Joseph M. Jauch, etc.), of which von Neumann was a pioneer.[8]

(iii) The third and last period is related to the groups of outsiders who started working again on FQM, consequently to the concrete alternative formulations of QM in terms of realistic *hidden variable* theories (of which the initiator was David Bohm in 1952), and the predictions of Bell's inequalities (Bell, 1964), which allowed to experimentally discriminate between local realistic theories and QM. Actually, these results remained almost completely neglected until the 1970s, and became rather popular only in the 1980s. It is on the side of this generation of dissidents, who once again were challenging the orthodox interpretation of QM, that Popper re-entered the scene of quantum physics. The rest of the present paper is devoted to this third period, which opened in 1980.

### 3. Popper's return to the debate on QM and his relationship with Jean-Pierre Vigier.

As argued previously, Popper developed very close friendships with several prominent physicists and, as time passed by, his relationships extended to the new generation of 'uneasy' physicists, concerned with foundations of quantum mechanics. We have to remember that, besides a few exceptions, physics in 1960s and 1970s was conducted within a completely pragmatic framework. Heavily influenced by the Cold War (military) demands, research in physics was focused mainly on those fields which could have had an immediate practical application (this gave birth to the notorious expression "shut up and calculate!"), whereas genuine theoretical research was completely dominated by high energy physics. Philosophy and physics had never been more far apart.

On the other hand, a small "quantum subculture" (Clauser, 2002) strived throughout the seventies to promote a revival of FQM, facing the reluctance and even the hostility of the majority of the physicists. This has been recently reconstructed by Olival Freire Jr. (besides others), in a comprehensive research published in a book with the significant title *Quantum Dissidents* (Freire, 2014). Such an open "opposition" of the physics community was felt also by John Bell himself, who, after a talk he gave at the ETH in Zürich, proudly affirmed: "I could beat them", referring to the audience mainly composed of particle physicists.[9]

Nonetheless, the following decade opened with the realisation of the first reliable experimental tests of Bell's inequalities (Aspect, Grangier and Roger, 1981), which inevitably represented a turning point in the acceptance of research on FQM, and finally drew the attention of a broader public to this issue. Although it is ought to be recalled that FQM, was to became a mainstream discipline only several years later with the advent of quantum information theory, it is remarkable that the experimental realisation of the neglected Bell's inequalities had immediately a certain resonance. The theoretical debate, previously considered merely philosophical, was suddenly turning into the interpretation of concrete outcomes of optical experiments, thus 'actual' physics. It is of some interest that this also had some popular resonance, as some notable newspapers published surveys devoted to Bell's inequalities and their consequence on local realism.

---

[8] A modern analysis of Popper's work in the field of logic of QM, has been recently proposed by Dalla Chiara and Giuntini (2006).
[9] The episode is recalled by Prof. Reinhold Bertlmann, a friend and collaborator of Bell in CERN. He also remembers: "in older talks I participated to, I could feel the great tension between the, say, conservative audience and John Bell. So it was for me like one man fighting against all the others". (Reinhold Bertlmann, interviewed by the author on 17/10/2016).

This was the case in *The Times* on August 28, 1981 and in the French *Le Monde* (which directly involved Popper, see section 4.2).

It was in the same period that Popper became once again closer to the research in physics, and started having direct collaborations with some of the protagonists of the revival that FQM was experiencing. About this, in a recent paper, Shields maintains that

> Popper's solitary efforts to offer a different view of quantum phenomena acquire a different status in the early 1980's. By this time, he had acquired colleagues in the theoretical physics community, one of whom, French physicist Jean-Pierre Vigier, had an international reputation (Shields, 2012, p. 4)

In what follows, it is shown how these acquaintances have been established – in some cases, like Vigier's, actually much before early 1980s – and it is described in detail the influence that Popper had on (a certain part of) the physics community in those years.

Something which contributed to the renewed interest of Popper towards QM, is undoubtedly the cultural milieu at the University of London in the late sixties and seventies, when very lively discussions took place around the Department of Philosophy of Science at Chelsea College, carried out by Heinz Post and Micheal Redhead. Basil Hiley,[10] the closest collaborator of David Bohm at Birkbeck College, remembers those days as very exciting, when together with Bohm and Popper attending the seminars on quantum theory. Also Vigier, who often used to go to London to hold discussions with Bohm and Hiley himself, gave occasional seminars which invariably Popper attended. On the other hand, some physicists from Birkbeck College use to follow the seminars held at the Department of Logic and Scientific Method in the London School of Economics. Among the participants were Imre Lakatos, Paul Feyerabend and Popper himself, who was professor at that institution.

### 3.1 The Garuccio, Popper and Vigier's paper

The real protagonist, and principal cause, of Popper's new effort into the world of research in physics was beyond the shadow of a doubt Jean-Pierre Vigier. Although it is difficult to precisely reconstruct when the contact between the two of them was firstly established, we can hypothesise that this happened during one of the abovementioned seminars, and in virtue of the common acquaintance with Bohm, who was in touch with Popper at least since 1959 and had co-authored a paper with Vigier in 1954 (Bohm and Vigier, 1954).[11] In this regard, in April 1957, Bohm and Vigier attended a conference of the Colston Research Society, *Observation and Interpretation: a Symposium of Philosophers and Scientists*, held in Bristol. Although Popper did not attend, he sent a paper and therefore he also probably read the other contributions in the proceedings (Körner, 1957), coming to know of Bohm's and Vigier's critique of QM already at that time.[12] In any case, the first documented conversation between Popper and Vigier dates back to as early as May 1966, when they met at the *First International Colloquium on Physics, Logic and History* held at the University of Denver, Colorado. There, both Vigier and Popper presented contributions, and during the discussion, Popper opened his speech saying "one word to my friend Vigier" (PA, 82/28),[13] which leaves no doubts on the fact that the two of them had interacted before that time.

---

10 I am most grateful to Prof. Basil Hiley for sharing this testimony in a personal communication, on October 10, 2016.

11 The first recorded correspondence between Popper and Bohm dates back to September 9, 1959 (PA, 278/2). While the quoted co-signed paper of 1954 was the result of a collaboration that Vigier had with Bohm, when he visited the latter in Brazil (see Holland, 2004).

12 I am thankful to Prof. Olival Freire Jr. for having pointed out to me this possible early contact between Popper and Vigier (communication to the author on January 29, 2017).

13 In Popper's archive a typewritten version of the discussion report the quoted sentence. Nevertheless it seems that in the published proceedings (Yourgrau, 1966) the sentence has been erased, probably because the recorded discussions underwent a more formal revision before publication.

Nevertheless, their documented correspondence does not start before June 29, 1980,[14] when Popper wrote a letter to Vigier, of which the content has a tremendous historical value, for two different reasons.

Firstly, in this document Popper wrote to his friend Vigier that he had not had "two happier afternoons for many years" and that Vigier's "work is most exciting and encouraging" (PA, 358/25). In that period, Vigier had just started a collaboration with the theoretical physics group of the University of Bari (Italy),[15] of which the head was Franco Selleri. In particular, Augusto Garuccio (Selleri's pupil and collaborator) spent a yearlong collaboration in Paris with Vigier, in 1980. They worked on a modification of the Mandel-Pfleegor experiment (Garuccio and Vigier, 1980), which could have experimentally revealed the existence of the *De Broglie Waves*. Vigier's collaborators recall that he went to London to discuss the subject with Popper and was eventually able to get him involved.[16] The mentioned two happy afternoons are most probably the testimony to Vigier's endeavour to recruit Popper.[17] In fact, in the same letter (29/06/1980), Popper was suggesting several improvements to an experiment proposed by Vigier (which, as one can easily deduce from the letter, is his modified version of the Mandel-Pfleegor experiment). In the subsequent reply on September 15, 1980, Vigier enthusiastically wrote to Popper:

> your proposal to improve on our laser experiment is excellent. In fact I am thinking of including it into a more detailed discussion [...]. Would you do us the pleasure of co-signing this improved version with Garuccio and myself? (PA, 358/25)

Popper answered that he did "not deserve it" (letter to Vigier of 07/10/1980, PA, 358/25), but Vigier did not give up, stating: "I feel your suggestion [...] is so important that I would not dare publish [*sic*] it without your signature" and he concluded the letter affirming: "it is a great pleasure (and honor) to fight for the old Einstein on your side" (15/10/1980, PA, 487/11).

So Popper, under the persistence of Vigier's proposals, eventually accepted, as proved by the two publications which respectively appeared on *Epistemological Letters*, in July 1981 (Garuccio, Popper and Vigier, 1981a) and on the more notable *Physics Letters A*, in December (Garuccio, Popper and Vigier, 1981b).[18] The aim of the proposed experiment - and this can easily explain Popper's interest in collaborating with Vigier - was to demonstrate "conflicting measurable predictions of the Copenhagen and statistical interpretations of quantum mechanics" (Garuccio, Popper and Vigier, 1981b). The authors continued stating that "their detection would help to choose between the antagonistic positions of Bohr and Einstein in the Bohr-Einstein controversy". It is worth mentioning that before appearing, the paper "provoked quite a lively discussion, not only with the Phys-Rev[19] referee's [*sic*] (one of whom was Shimony) but also with Bell and Aspect and others" (Letter from Vigier to Popper on 09/04/1981, PA, 487/11). Indeed, in the latter published article, the authors acknowledged Leonard Mandel (who co-authored the original experiment modified by

---

14 As already recalled, Vigier's documents are possibly irretrievable (see footnote 6).

15 The genesis of this collaboration has been reconstructed by Baracca, Bergia and Del Santo (2016), section 13.2.2.

16 I am grateful to Prof. Agusto Garuccio and Prof. Nicola Cufaro Petroni, who shared with me their recollections on the relationships between Popper, Vigier and the group in Bari, in a joint personal communication on October 11, 2016.

17 From Vigier's letter to Popper on 15/09/1980 we know that they met in that occasion in London (PA, 358/25), while from another letter to Popper dated 15/10/1980 (PA, 487/11), we know that the two had their discussion after a talk that Vigier gave. This can have happened in one of the already remembered seminars at Chelsea College or at the London School of Economics.

18 It is remarkable that Garuccio, although co-author of the work, did not meet Popper at that time (Augusto Garuccio, personal communication to the author on 12/10/2016), and neither did any other Italian physicist working on FQM until 1983 (see section 4.3).

19 This paper was initially submitted to *Phys. Rev. Lett.* but rejected (letter from the editor in date 01/04/1981, PA, 338/15) because answering the referee's requests would have enlarged too much the contents, so that the editor suggested to submit it to *Phys. Rev.* And so did Vigier, but then he "sent it to *Physics Letters* because [he] felt one of the Phys. Rev. Referee's was unconvictable for unscientific reasons" (Letter by Vigier to Popper on 09/04/1981, PA, 487/11).

Garuccio and Vigier), John Bell, Alain Aspect and Basil Hiley for the helpful discussions, and also Franco Selleri, for his first proposal to detect de Broglie Waves.[20]

Moreover, it is noteworthy that a conference in honour of Karl Popper took place in Cerisy-la-Salle (France) on 1st-11th of July 1981, approximately in the same days when the first paper co-signed by Garuccio, Vigier and Popper appeared. There, Vigier gave a talk with the title "Le réalisme de la mécanique quantique" (published in the proceedings as "Popper et le débat Bohr-Einstein"), where he also described the state-of-the-art experiments on the subject, including his own collaboration with Popper (Bouveresse-Quilliot, 1989, p.297), but he did not mention Popper's EPR-like experiment, which had not been published yet at that time (see next section).[21]

Although recent historiographical literature on Popper's contribution to quantum physics (Freire, 2004; Shields, 2012) only mentions the latter of the two papers (which appeared in *Physics Letters*), probably because it was rightly considered more influential, one must emphasise that the first one has a certain relevance, not only because appeared five months before, but also because it actually gathered the attention of some physicists active in the field of FQM (see footnote 25 for an insight on the journal *Epistemological Letters*). Indeed, the physicist Olivier Costa de Beauregard, from the *Fondation Louis de Broglie* in Paris, immediately submitted a critical reply to Garuccio, Popper and Vigier, published in the very same issue of *Epistemological Letters*. So it is still correct, with Shields (2012, p. 5), that the "first to respond was French physicist O. Costa de Beauregard, who offered in May of 1982 a brief letter entitled *Disagreement with Garuccio, Popper and Vigier*" (Costa de Beauregard, 1982), but his first response actually had appeared already in July 1981. Moreover, although *Physics Letters A* published Costa de Beauregard's paper in the issue of the 10th of May 1982, this had been received only on January 30, whilst a paper was submitted by Mandel in the same journal, on January 27, 1982, yet published only on May 31. There, Mandel claimed that there are "some difficulties with a modified version of the Pfleegor-Mandel experiment recently proposed by Garuccio, Popper and Vigier" (Mandel, 1982). While Costa de Beauregard directly claimed that "the way Garuccio, Popper and Vigier have recently used quantum mechanics […] does not only conflict their own SIQM [Statistical Interpretation of QM], but also the way CIQM should be used properly." (Costa de Beauregard, 1982).

For what concerns Vigier and Garuccio, they published a new paper (in July 1982), where the collaboration with Popper was replaced by that with the Italian experimentalist Vittorio Rapisarda (see Baracca, Bergia and Del Santo, 2016, session 13.2.2), and the experimental apparatus was improved. A personal communication from Popper is reported in the references, as a witness that Popper remained involved in this research (Garuccio, Rapisarda and Vigier, 1982, reference 4).

### *3.2 The genesis of Popper's experiment*
Coming back to the letter to Vigier dated June 6, 1980, the second and main reason for which it has a pivotal historiographical importance, is that Popper formulated there for the very first time his simplified version of the EPR experiment (hereinafter referred to as just Popper's Experiment, PE), which was to be firstly published only as late as in 1982 (Popper, 1982a; 1982b). The subject later on received great deal of attention and divided the community of physicists on more the one occasion. In what follows, it is shown how Popper's important and controversial thought experiment originated.

---

20 As a matter of fact Selleri started his career on FQM in 1969, with a short work on the reality of empty waves (see Baracca, Bergia and Del Santo, 2016, section 3), then followed further this line of research also publishing a long paper which integrated Garuccio, Vigier and Popper's work (Selleri, 1982).
21 The philosopher of science trained as a physicist, Michel Paty, delivered at the conference in Cerisy-la-Salle the paper "Popper et le débat quantique".

To start with, it is of great interest to reproduce the first formulation, never published before, of PE (from a letter to Vigier on 09/06/1980, PA, 358/25): [22]

> Now after the meeting and encouraged by our conversation, I had an idea for the variant of EPR which is terribly simple and which will, I think, come out against Copenhagen. And today I simplified it further:
>
> <u>A Simple Version of the EPR Experiment.</u>
>
> (1) We start with an arrangement with a source <u>S</u> (positronium, say) in the centre and detectors and coincidence counters to the right and to the left, along the y-axis, say. <u>No polarisers are needed</u>; instead we have screens with variable (parallel) slits to 'measure' the x-position $q_{xA}$ and $q_{xB}$. The width of the slit is $\Delta q$, which can be varied.
>
> (2) We find positions for the slits A and B such that we get satisfactory coincidences for the width $\Delta q_{xA} = \Delta q_{xB}$.
>
> (3) We have a whole battery of detectors behind the slits which we make small so that, owing to $\Delta p \Delta q \geq h$, we get a wide scatter of momentum. We can make a statistics of the scatter angles.
>
> (4) We now open wide the slit at B: $\Delta q_{xB} \gg \Delta q_{xA}$.
>
> (5) Argument: By measuring the position of A, $q_{xA}$, with the precision of $\Delta q_{xA}$, we indirectly measure the position of B with approximately the same precision. From the point of view of the Copenhagen interpretation (as defended by Bohr in the Schilpp volume on Einstein) the momentum of B ought to be smeared, which means that the momenta of a beam $B_s$ (= B-particles) ought to scatter as before. Thus, we ought to get the same scatter statistics, even though the slit at B is now wide open.
>
> <u>This can be tested</u>. If, as to be expected, there is little scatter, less than before, then the Copenhagen Interpretation is refuted.
>
> Many many thanks and kindest regards,
>
> *(6) It seems that this simple experiment shows that the Copenhagen interpretation actually contradicts Quantum Mechanics. For Q.M. says that a change of the momentum $p_{xB}$ of the particle B is possible only if there is a body (such as the screen) which can receive the momentum: the <u>knowledge</u> thus has to have the same effect as the screen.*
>
> *It is to be expected that the CIQM will be refuted and QM remains correct.*

This extremely valuable document, predates Popper's experiment by two years. What is significant - besides that the experiment remained basically unchanged throughout the upcoming years - is that Popper could eventually formulate an experiment capable (or at least he claimed so) of discriminating between the (subjectivist) CIQM and a realist interpretation. Also Einstein in his original EPR paper (Einstein, Podolsky and Rosen, 1935) was aimed to show the incompatible nature of CIQM with the commonsense hypotheses of local realism (for a complete theory). It is doubtless that Popper was strongly influenced by Einstein since as early as 1919, as he recalled in his intellectual autobiography

> Einstein "became a dominant influence on my thinking—in the long run perhaps the most important influence of all" (Popper 1976, p. 37).

---

22 In the folder containing these correspondence are collected several drafts of this letter, both hand- and typewritten. Nevertheless, in this final version (which we can presume it is the one that Vigier received), the sixth and last point of the list is a handwritten addition which is reproduced here in italic, to highlight this difference. The underlined phrases are in the original letter.

It is also well known that Popper had been working on thought experiments, aiming at falsifying CIQM – and specifically Heisenberg's Uncertainty Principle – since his mistaken paper of 1934 (see section 2). Although Popper remained forever ashamed of this episode, "one can even speculate that Einstein was influenced by Popper" in deriving his EPR experiment one year later (Redhead, 1995).[23] In any case, we can surely claim that Popper never abandoned the idea of proposing a crucial experiment to empirically test different possible interpretations of QM. And, as I showed, this happened by means of the new stimulus provided by the meeting with Vigier (which one can suppose to have taken place a few days before their correspondence of 09/06/1980). My hypothesis here is that Vigier's proposal of experimentally testing the statistical (realist) interpretation of QM breathed new life into Popper's project of empirically falsifying CIQM. On the other hand, one has to recall that Garuccio and Vigier, had already submitted, on March 19, 1980, a paper on their own on the same subject (Garuccio and Vigier, 1980). In conclusion, Popper only gave minor, and actually technical,[24] contribution to an already existing research project carried out by Vigier and Garuccio in Paris, and it was just because of Vigier's resolute initiative that he got involved. At the same time the concrete possibility of testing different interpretations of QM, claimed by Vigier, stimulated Popper to formulate his new thought experiment, though based on an old personal struggle.

So my reconstruction seems to solve the 'oddity' that Shields pointed out:

> Oddly, this experiment [...] is not at all the same as in the paper with Vigier and Garuccio, though the underlying purpose is the same. (Shields, 2012, p. 5)

But in the perspective that I have exposed, it seems rather natural that the two works exhibit totally different physical apparatus and underlying mechanisms to achieve the same aim, because they have been hypothesised independently by Popper and Garuccio-Vigier, respectively.

Vigier answered to Popper that his "proposed experiment is very nice" (letter to Popper on 15/09/1980, PA, 358/25). Already since the very beginning, it is quite clear that he was not enthusiastic about PE, if one compares it with how Vigier reacted to Popper's advice concerning the modified Mandel-Pfleegor experiment. Nonetheless, he encouraged Popper: "submit it for publication in the Epistemological Letters to test our opponent's reactions... and then send it to the Physical Review Letters."[25] But Popper did not follow this advice, and preferred to integrate this result in his upcoming book on FQM (see section 4.1).

**4. Popper's last entry in the physics community**

*4.1 Publication of the experiment and the weak response of the physicists*

As described in the last sections, Popper made his (new) debut into the physics community in 1981, by virtue of Vigier's initiative. But at that time he had already spent several decades fighting against CIQM and its subjectivism. Popper defended realism not only as a possible interpretation to be given to physical phenomena, but rather as a necessary foundation for physics itself, as he was to summarise a few years later:

---

23 A short, but well documented historical overview of the possible influence of Popper's thought experiment of 1934 on Einstein, has been reconstructed directly by Popper (1982a, p.15, footnote 20).

24 In the quoted letter of June 9, 1980, Popper proposed a series of physical improvement of the two-laser experiment, and not even a single philosophical concept is mentioned there.

25 The reader should notice that this double-submission process is the very same that the paper of Garuccio, Popper and Vigier underwent. In fact, the journal *Epistemological Letters*, defined by another of the protagonists of FQM, John Clauser, an "'underground' newspaper, whose circulation was limited to members of a 'quantum-subculture'" (Clauser, 2002), was the testing ground for every physicists working on FQM.

> I am a realist. Indeed, I am not only a realist but a metaphysical realist, as I want to admit at once. That is, my realism is not based on physics; but physics, I think, is based on realism.
> (Popper, 1985, p. 3)

Such an endeavour led Popper to systematise his positions and opinions in a comprehensive work on quantum theory, which was finally published, after many delays, in one of the three volumes of his *Postscript to the Logic of Scientific Discovery*.[26] In fact, the third volume, entitled *Quantum Theory and the Schism in Physics*, is a 200-page treatise completely devoted to a careful defence of "a realistic and commonsense interpretation of quantum theory" (this is the title of the preface in Popper, 1982a). Popper worked on this book for about thirty years, yet at the beginning of the 1980s the publication was finally on course, so he probably found it natural to include his new EPR-like experiment there (Popper, 1982a, pp. 27-30), instead of submitting it for publication in the form of a research paper on a physics journal, as Vigier had suggested to him (see section 3.2).

The only remarkable new element in the published experiment is the additional interpretation that Popper gave to it and which he maintained ever since. In fact in the published work, he claimed that his experiment, besides the main result of putting CIQM to the test, could have discriminated between Lorentz's and Einstein's interpretations of relativity, in the case of a result indicative of non-local interactions (Popper, 1982a, pp. 29-30).

Remnants of Popper's handwritten notes (PA, 223/17), show a draft of what most probably were to become a considerable part of sections VI-IX of the third volume of his *Postscript* (Popper, 1982a). In the document one can find two drawings realised by Popper (here reproduced in Fig. 1) which were to become figures 1 and 2 of Popper's *Quantum Theory and the Schism in Physics* (here reproduced in Fig. 2).

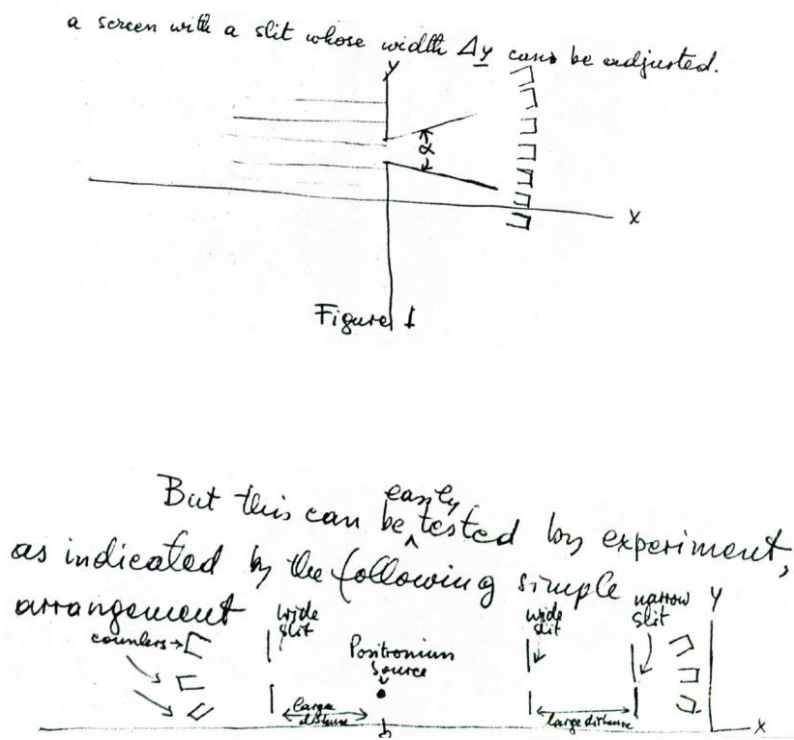

Figure 1. *Unpublished original drawings made by Popper in 1980 (PA, 223/17) to graphically explain his experiment. They were to became the figures of the third volume of his Postscript, two years later (cf. Fig. 2 ).*

---

26 The main text of these three volumes was written as early as in 1950s, but the introductory part in which PE firstly appeared belongs to the complete edition of 1982. The story of the difficulties in the publication of these books are described by Shields (2012, section 3).

Although undated, these notes contain elements which allow to indirectly date them to 1980,[27] therefore in the very same year of the conception of PE (see section 3.2).

A further consideration on the genesis of PE, can be inferred by other information given by Popper in the quoted letter to Vigier of 09/06/1980. There Popper cited the position of "Bohr in the Schilpp volume on Einstein" (Schilpp, 1949). This outstanding book contains indeed an essay authored by Bohr, on his controversy with Einstein (pp. 199 - 243), where a slit-diffraction example is used to justify Heisenberg's relations. It is crystal clear that Popper had studied the Copenhagen interpretation (also) on this volume, and in particular his idea - exploited in PE - of using narrow slits to spread out the momentum indeterminacy in order to falsify Heisenberg's principle, has probably been stimulated directly by the examples displayed by Bohr in this book.[28]

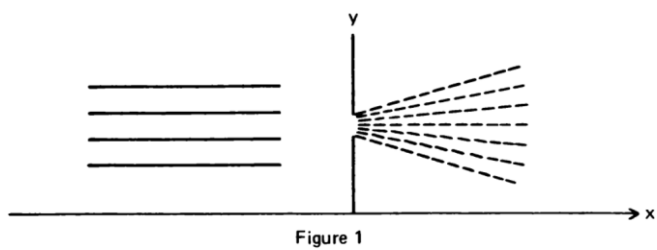

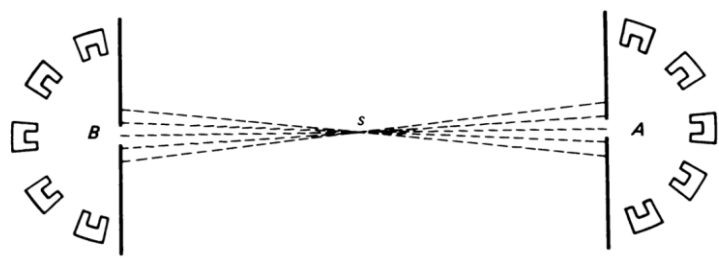

Figure 2. *Reproduction of figures 1 and 2 of Popper's Postscript to the Logic of Scientific Research (1982a, p.17 and p. 28, respectively). They explain the spreading in momentum due to the presence of a narrow slit (above) and the proposed setting of the source and of the detectors (below) in Popper's EPR-like experiment.*
With permission of University of Klagenfurt / Karl Popper Library. All rights reserved

Popper's experiment appeared a second time, again in 1982, on a three-page contribution printed in a collective volume in honour of Carl Friedrich von Weizsäcker's 70th birthday (Popper, 1982b). One can stress that in this collection of papers, entitled *Physik, Philosophie und Politik*, Popper was contributing in the session dedicated to physics, at the side of Lutz Castell, David Finkelstein and Asim Barut, all theoretical physicists. Nevertheless they were not among the ones uneasy towards QM, so that also this "Proposal for a

---

27 In the document, Popper stated that 55 years were passed from the formulation of QM made by Heisenberg (surely referring to the so-called matrix mechanics of 1925), and 45 years from EPR publication (1935).

28 Popper repeated in several occasions that the first (mistaken) experiment he invented in 1934 was the 'predecessor' of the one under discussion. Nevertheless, I deem it necessary to clarify that the proposal of 1934 was based upon a different physical mechanism (i.e. the interaction of beams of photons and matter particles respectively) although it was as well aiming at falsifying Heisenberg's relations. So, even thought it also made use of a slit to collimate one of the beams, the increased diffraction due to the Heisenberg principle, which is explicitly outlined by Bohr (Shlipp, 1949), was not exploited in the first *Gedankenexperiment*.

Simplified New Variant of the Experiment of Einstein, Podolsky and Rosen" probably did not circulate within the community interested in FQM.

In conclusion, it is more than likely that just very few physicists read Popper's experiment immediately after its (double) publication in 1982, and, as far as I could reconstruct, none of those who were at the same time struggling to preserve realism in QM also read. Thus Popper's book on QM remained confined to the domain of philosophy, where not many scientists are interested to venture, and even less so at that time. Even the most open-minded physicists at the front line for realism, like Selleri and Vigier (who had been the very first one to know about PE), did not read *Quantum Theory and the Schism in Physics* until 1983, when Popper gave them the book as a present.[29]

Therefore Popper would have waited still another year before his experiment aroused the first interests and criticisms.

### *4.2 Foundations of quantum physics break into newspapers: the case of "Le Monde"*

As an additional evidence of Popper's full involvement in the debate on FQM, which since the 1980s had a growing audience, it is interesting to consider the role that the French press played in the popularisation of FQM.

*Le Monde,* one of the most respected and read French newspapers, had already a tradition of articles both on philosophy of science and on physics, and in particular it had devoted some attention to the wave mechanics, surely because of the French most distinguished figure in quantum physics, the Nobel laureate Louis de Broglie. But, at the end of the 1970s, *Le Monde* started printing on its pages open debates on the (also less orthodox) interpretations of the new results of FQM. On October 24, 1979 the journalist Michel Kajman entered an article reporting on a conference held in the same month in Córdoba (Spain), and entitled *Science and Consciousness.*[30] This was devoted to the "so-called 'parapsychological' phenomena (remote viewing, telepathy, psychokinesis) [which] had became object of study for a certain number of physicists ".[31] What is more remarkable is that in the following pages there were published a series of articles collecting the opinions of some of the most famous dissidents of quantum theory, namely David Bohm, Jean-Pierre Vigier, Franco Selleri, Olivier Costa De Beauregard and the American Harold Puthoff and Fritjof Capra (the latter two were playing a central role in the nascent counterculture around quantum physics in US, see footnote 30). Also an article entitled "Le Paradoxe E.P.R.", by Maurice Arvonny, was published on the same issue of the newspaper; while articles authored by Aspect and then by Vigier themselves appeared on *Le Monde* on March 14, 1980 and April 4, 1981, respectively.

Aspect's experiments were receiving increasing attention and, as the results were refined, physicists working on FQM had to face the real problem of interpreting them. The ideas of Popper, Vigier and Selleri, but also those of Bell, Bohm and in general the new generation of 'quantum dissidents', had a solid common ground based upon realism, but locality there had always been an element of discordance. And what better occasion to give the coup de grace to the already fragmented groups dedicated to FQM, than that of isolate them on the basis of the problem of (non-)locality, by virtue of Aspect's novel results.

---

29 In a letter dated January 12, 1983 to Vigier (PA, 358/25), Popper enclosed copies of both *Quantum Theory and the Schism in Physics* (stating that "only pp. 27 to 30 are essential", namely those containing PE) and his contribution to Weizsäcker's 70[th] birthday volume. While Selleri (who, we must remark, did not have contacts with Popper before 1983), wrote to Popper on December 12, 1983: "I am reading [...]Quantum Theory and the Schism in Physics (your gift)." (PA, 348/21).

30 The conference, held on 1st-5th of October 1979, was organised by Radio *France Culture* (The proceedings have been printed in French, in a collective volume: *Science et Conscience. Les Deux Lectures de l'Univers.* 1980. Stock, Paris) and was devoted to the alternative scenarios that the recent results confirming non-locality seemed to open. We have to stress that this event was rather peculiar in the European milieu, and would deserve further attention. In fact, in Europe the New Age attitude and the alleged parapsychological applications of non-locality did not typically enter research in physics, but it was on the contrary extremely important in the development of a scientific counterculture in the US, as reconstructed by David Kaiser (2011) in his book *How the Hippies Saved Physics*. Harold Puthoff and Fritjof Capra, who also contributed into *Le Monde*'s pages, were among the exponents of such a ''hippie' climate into physics in US.

31 Michel Kajman, Nouvelles Frontières Et Vieux Débats À Cordoue, *Le Monde*, 24/10/1979.

So a first article signed by Claude Serraute appeared on Le Monde on December 12, 1982 with title "Et si Dieu jouait aux dés?" (What if God Played Dice?), and was followed in rapid succession (15/12/1982) by a second one, entitled "Dieu joue probablement aux dés" (God Probably Plays Dice) by Maurice Arvonny. This immediately gave rise to a counteroffensive of the most strenuous defenders of Einstein: Vigier among them. Worried by this attack, he wrote to Popper:

> Le Monde's paper contains two things
>
> a) A fair and objective account of Aspect's experiment and results. […]
>
> b) A theoretical analysis against Einstein... concluding that God probably plays dices. In other terms
>
> − realism is dead
>
> − causality also
>
> […] One can easily imagine how this result will be exploited in various circles all over the world. A certain number of people have thus reacted including Kastler (the French Nobel prize in physics) and suggested to le Monde to make a double page on January 16th 1983. […] Contributions […] have been requested from
>
> yourself
>
> John Bell
>
> Bernard d'Espagnat
>
> Selleri
>
> and myself

Vigier also gave Popper some advice regarding the possible contents of the latter's contribution:

> recall the importance and history of E.P.R. Paradox (including your own contribution).
>
> […] If subsequent experiments confirm Aspect's results then one could drop determinism (as Einstein himself had suggested) but not necessary realism. (PA, 358/25)

Vigier's letter stimulates several considerations. The first and more general one, is that Aspect's result came as a sudden shock for most of the physicists who have been advocating Einstein's ideas for decades, and for those who caught hold of Bell's theorem aiming at finding limits to the validity of QM. Yet the experimental results were not only confirming QM, but they were also making more and more clear that some of the cornerstones between realism, locality (and determinism) should have been forever abandoned. Vigier, not without being reluctant, was ready to renounce locality and, already for some years, was working on a way to preserve relativity in case of the confirmation of non-local effects (see section 4.3).

Popper, who had been opposing determinism for many years (see e.g. Popper, 1950), claimed:

> The attribution to Einstein of the formula 'God does not play with dice' is a mistake. Admittedly, he was a strict determinist when I first visited him in 1950. […] But he gave this up.
>
> [...]Einstein was, in his last years, a realist, not a determinist.
>
> (letter to Vigier of 12/01/83, PA 358/25)

Whereas, about dropping locality, he had a totally different idea. Actually he had a different idea than anyone else about the interpretation that should have been given to Aspect's experiments. He did not believe that the original EPR experiment, based on position-momentum correlations, and its analogous version, formulated by Bohm (EPRB) in terms of spin (or photon polarisation), were physically equivalent therefore, it is conceivable that EPR and EPRB would lead, if experimentally tested, to very different results (letter from Popper to Vigier of 12/01/83, PA 358/25)

And, since Aspect's experiment was testing a Bell inequality based on EPRB, Popper maintained:

> Aspect's most exciting experiment is an <u>experimentum crucis</u>, a crucial experiment.
> But I suggest that we ought to consider whether it is crucial
>   (a) between local action and non-local action, OR
>   (b) between the statistical interpretation and the Copenhagen Interpretation.
> […] So I suggest that Aspect's crucial result <u>may</u> allow for a different interpretation. […] If Q.M. is a local theory, then Aspect's experiment may show, by showing its correctness, only that Copenhagen interpretation is mistaken.
> […] it is, for this reason, of interest to try out the version of EPR which I have proposed twice. (letter to Vigier of 12/01/83, PA 358/25)

This is a perfectly tenable and logically consistent position, but it presupposes that Bell's inequalities are not generally valid, as in fact Popper was maintaining (see. Section 5.2).

Another remarkable aspect is the one pointed out by the list of the requested contributors by *Le Monde*: Popper's name is among name is among those of the most prominent physicists in the field of FQM at the time. So his expertise in foundations of quantum physics was very much recognised, and as Vigier stressed in the same letter: "many people are anxious to know how you will react to the new situation created by Aspect's results"(PA, 358/25).[32]

Most surprisingly, it seems that none of the mentioned contributions appeared on the pages of *Le Monde*, neither on the expected date in January 1983, nor later.

### *4.3 The Bari workshop on Open Questions in Quantum Physics of 1983*

On May 4th-7th 1983, an international workshop, organised by Franco Selleri, was held in Bari with the title *Open Questions in Quantum Physics*. Selleri himself remembered in an interview twenty years later that the

> conference was organized at the time to attract Karl Popper whom I understood then to be a very important philosopher for us, because Karl was very critical of the Copenhagen approach. (Freire, 2003)

This marks the beginning of the contacts between Selleri and Popper. Vigier was again the intermediary and likely also the promoter of the event, since Selleri admitted that only afterwards he understood the importance of Popper for his cause.[33]

---

32 It ought to be stressed that this can probably be regarded as the period of highest mutual appreciation and affinity between Popper and Vigier. The latter was making a concrete effort to keep Popper into the international scene of quantum physics, as proved by the kind proposal that Vigier made to him: "I have promised *Le Monde* that I would personally translate your text I propose to come to England[...] and do the translation under your eyes [...]. This I think is justified by its importance" (letter to Popper, undated, PA, 358/25). Popper answered, in the mentioned letter of 12/01/1983, with a 9-page handwritten letter where 14 key points were listed as the backbone of his contribution supposed to appear into *Le Monde*. The mutual blind faith between the two of them is attested by the words of Popper: "I authorise you to translate <u>everything</u> from my letter, from my book [Popper, 1982a], or from the Sonderdruck [Popper, 1982b] for Le Monde; to omit anything you like; and to re-arrange it as you think best!" (PA, 358/25).

33 Garuccio and Cufaro Petroni remember that the contact between the group of Selleri in Bari (of which they were part) and Popper was surely mediated by Vigier (personal communication on October 11, 2016). Moreover a short undated note, undoubtedly written by Popper to Vigier (PA,

After the first invitation letter from Selleri to Popper on March 5, 1983 (PA, 247/18), the following correspondence between the two shows the great importance that the Bari conference was to have for Popper. In fact, this was the long-awaited occasion to introduce his experiment to a wider audience of physicists working on FQM, and try to convince them of its validity. This can be seen already in another letter which Selleri had sent to Popper before the conference, where one can read:

> I read with great interest your proposed experiment and I think that it can be done. Some experimentalists will be in Bari during the meeting and perhaps we will find somebody who does it. (Selleri to Popper on 08/04/1983, PA, 247/18)

The conference opened with the talk given by Popper, "Realism in Quantum Mechanics and a New Version of the EPR Experiment". There Popper explained, in a clear and detailed exposition (reported in the proceedings as a 23-page contribution: Popper, 1985), his positions concerning QM. I agree with Shields (2012, p. 7) when he states that the Bari conference represents for Popper "the culmination of his thinking on quantum theory". In fact, by that time (most probably due to the influence of the collaboration with Vigier) Popper had had developed new perspectives on his book from two years before (Popper, 1982a). For instance, he had completely embraced the theory put forward by de Broglie-Bohm-Vigier and Selleri of the wave-particle dualism, in terms of real empty waves[34]

> I as a realist assert, with Louis de Broglie, that particles, which are carriers of energy, are always accompanied by waves, while waves are perhaps not always accompanied by particles: *there may be empty waves.*(Popper, 1985, p. 5)

But, as already recalled, the actual reason for which Popper deemed the workshop in Bari to be of central importance, was the possibility of convincing some of the attending physicists to carry out his experiment. This is made clear by Popper's own assertion: "I plead here only that my experiment should be conducted by somebody." (Popper, 1985, p. 10).

The presentation of PE in Bari was enlarged with a more careful account of its hypothetical results, although the form of the experiment had remained substantially unchanged since its very first formulation (see section 3.2), Popper claimed that his experiment could have had, if performed, one of the following possible outcomes:

> (1) Perhaps the particle will go on and hit one of the central counters, while its twin-brother on the right, after passing the slit, shows a more radical scatter. [...] So the first possibility is that our *knowledge* has no physical effect. [...]"
> (2) [...] the particles going to the left will scatter (like those on the right) as a result of our knowledge of their position. This, I assert, is what Bohr and Heisenberg were committed to assert [...]"(Popper, 1985, p. 9).

Popper, advocating the first outcome, highlighted that the outcome of the second scenario depended on whether the particles would have undergone uncorrelated scatters or otherwise. He also reported that the

---

247/18) states: "I got a letter from Bari and I lost it! Please post the enclosed to the man in Bari (I do not know his name and address!)". In the same letter one also learns that Vigier would have gone to London expressly to travel together with Popper to Bari. So, at this stage, Vigier was effectively Popper's reference.

34 I am grateful to Prof. Gino Tarozzi, who pointed out to me (personal communication to the author on 02/12/2016) that this position was not yet developed in the *Postscript* (Popper, 1982a), where Popper still maintained a corpuscular ontology in which the modulus squared of wave function describes statistical ensembles of particles.

latter "was the position taken up by Jean-Pierre Vigier when we discussed my experiment" (Popper, 1985, p. 9). The fact that Vigier expected instantaneous measurable correlations was the major element of disagreement between Popper and Vigier.[35] The French physicist was in fact concerned with the problem of locality at least since 1979, and, as a strong opponent of the CIQM, he seriously struggled to find an interpretation capable of coherently maintaining both realism and the increasingly improved evidences of non-locality, due to Aspect's experiments. As early as 1980, Vigier suggested Popper improve his EPR-like experiment (just after its very first formulation), providing devices similar to those used by Aspect, "to ensure superluminal interactions [...] (if they exist of course)". (15/09/1980, PA, 358/25). The doubt of the existence of superluminal signals matured in Vigier in the following years, and, when he shared with Popper a confidential preprint of the early Aspect's result, his uneasiness became manifest: "we must now face the monster of non-locality", he affirmed (Letter to Popper on 09/04/1981, PA, 478/11). The solution he developed is summarised in the following statement, which he wrote to Popper:

> Action-at-a-distance <u>can be causal provided certain covariant constrains are satisfied</u>. […] This is the case of EPR situation. (letter from Vigier to Popper on 16/05/1981, PA, 487/9)

Vigier was very anxious to know Popper's opinion on this - as he personally stated in two consecutive letters to Popper on April 4 (PA, 478/11), and May 16, 1981 (PA, 478/9) – because he probably did not totally rely on his own solution. Nevertheless he submitted these new results for publication two months later (Vigier, 1982).

At beginning of January 1983, in a letter to Popper, Vigier showed a much sharper position towards the results of Aspect's experiment (obviously also because the results were corroborated at that time). In the letter he claimed: "the general opinion here (among the physicists including myself) is that [Aspect's] experiment proves non-locality, directly independently of any interpretation". Nevertheless he stated that this could in any case be interpreted "in a realist way, but in that case one must accept non-local (superluminal) interactions as a fact." (PA, 358/25). Popper considered this position untenable (see section 4.2 and Popper, 1985), and so did Selleri, who many years later recalled: "[Vigier] considered himself a nonlocal realist, a position whose motivation for me is still very difficult to understand. It is like realism with miracles, but then one can keep the Copenhagen approach, which is full of miracles" (Freire, 2003).

Coming back to Popper's talk in Bari, this stimulated a lively discussion, reported in the proceedings, among Popper, Marcello Cini, Francesco De Martini, Karl Kraus, Trevor W. Marshall, Helmut Rauch, C. Robinson, Franco Selleri, J. Six, Gino Tarozzi and Jean-Pierre Vigier.[36] The debate divided the audience, and was mainly focused on the problem of the actual feasibility of such an experiment. Vigier in this occasion openly opposed Popper's viewpoint. Six was the first to point out the problem of realising a source of collinear photons, of which the wavelength would have to be compatible with the width of the slits. De Martini, the experimentalist from Rome who was the most likely candidate to carry out PE,[37] in fact proposed to conduct a variant of it, which however did not convince Popper. Gino Tarozzi, trained as a philosopher of science but who had an enduring scientific collaboration with Selleri, expressed his admiration for the "great relevance, both physical and epistemological, of Professor Popper's experimental proposal". Also Selleri, although aware of the practical difficulties, advocated the realisation of PE.

---

35 The dispute between Popper and Vigier, never took the tone of a fight, rather was even jokey. Just a week after their meeting in Bari, Popper wrote to Vigier: "as to our bet: you may remember that I suggested that one can telegraph with superluminal velocity if you are right. [...] So I may still win our bet in which case it is part of the bet that you look again, <u>very carefully</u> at Aspect" (letter of 15/05/1983, PA, 358/25)

36 These are only the physicists who contributed to the discussion, but many other physicists attended the conference, among them the abovementioned Basil Hiley, Emilio Santos, Nicola Cufaro Petroni, Augusto Garuccio, and Leonard Mandel.

37 Selleri wrote to Popper the month before the conference in Bari (22/04/1983) that "some experimentalists are coming to see if they are able to do your experiments [sic], dr. De Martini from Rome is one of them." See also section 5.

## 5. Further involvement of Popper in the physics community in the 1980s

### *5.1 The debate on Popper's Experiment in the 1980s*

Bari's conference marked to some extent the starting point of the public involvement of Popper to the debate on FQM. As I have shown, until that time Popper had had some opportunities to actively participate in the new quantum debate (mostly thanks to Vigier), yet his activities in this field had not had a tremendous resonance. Now, by means of this 'official' introduction into the community, which occurred in Bari, Popper's contributions started to be seriously considered within the most important meetings devoted to discuss the new developments of FQM.

After the conference in Bari, Popper was enthusiastic both because of the amazing intellectual affinity he discovered to have with Selleri[38] and because of the possibility of finally seeing his experiment realised. Yet he did not trust the commitments of the experimentalist who attended the workshop, as he wrote to Selleri:

> It was a wonderful conference, probably the best I ever attended.[...]
> As to my experiment [...] I am 81 and therefore rather anxious to see it realized [...], I fear that that [*sic*] Francesco De Martini (whom I liked very much) will not carry out his plan to do this experiment: he is in love with the idea to do his far easier (but in my opinion non-equivalent) time reversal of my experiment (PA, 247/18)

Nonetheless, Selleri took the proposal so seriously that he did not only appoint himself to check whether De Martini intended to perform the experiment (01/06/1983, PA, 247/18), but he even visited his laboratory in Rome. However, Selleri "was deceived from his [De Martini's] way of doing physics", so that the perspective of a collaboration with De Martini vanished (Selleri to Popper on 10/09/1983, PA, 247/18).

Selleri, for his part, worked on the theoretical foundations of PE, and wrote to Popper (10/09/1983, PA, 247/18) that together with Garuccio they realised that a source producing collinear photons compatible with diffraction-slit experiments could have actually represented a problem. Moreover, in a conference held in Athens in 1984,[39] Selleri presented again PE, arguing that a suitable source would have been theoretically realisable, although there were "many physicists saying that [Popper's] experiment is in principle impossible because the collinearity requirement cannot in principle be satisfied" (Selleri to Popper on 05/10/1984, PA, 562/19). However, Selleri's last word on PE was rather negative. Together with the Canadian physicist Don Bedford, who spent a sabbatical in Bari, they concluded that PE "is impossible in principle [...] in all those cases in which the emitting source disappears in creating the 'collinear' pair" (letter from Selleri to Popper on 12/12/1984, PA, 562/19). We must stress that at that time there was no known technique which did not make use of 'disappearing' sources to create entangled photons. Selleri, who published this result a few months later (Bedford and Selleri, 1985), did not give any further consideration to PE throughout the following years, and he did not even mention it in a book he published in 1989, although a whole chapter is devoted to "Popper and modern physics"(Selleri, 1989, pp. 169-187).

---

38 Popper and Selleri immediately established a very close relationship, about which some original quotes may help to understand the influence that particularly Popper had on Selleri. Just after the workshop of Bari, Popper wrote to Selleri: "for me the best was meeting you and became your friend." (May 1983, PA, 247/18). And Selleri replied: "you left behind not only feelings of intellectual admiration but of friendship as well." (01/06/1983, PA, 247/18). Such was the beginning of a long-lasting correspondence, based on a deep reciprocal respect and admiration, both at an intellectual and a personal levels. They started reading each other's books, and in particular Popper liked very much (although with some critiques) Selleri's *Die Debatte um die Quantentheorie* (Selleri, 1983). Selleri answered: "that you like my book is for me highly awarding and pushes me continue this (hard) struggle" (07/12/1983, PA, 348/21), and asked Popper to write an introduction for the upcoming translated editions of the book (Selleri to Popper on 05/10/1984, PA, 562/19), which it happened for the Spanish, Greek and French versions (Popper to Selleri, 28/10/1984, PA, 253/1). Whereas, after Selleri read Popper's *Postscript* (Popper, 1982a), he showed the following great appreciation: "only after meeting you I really understood why the cultural spaces for our activities are so much broader than they were in the fifties (or in the thirties). It is your merit, in a considerable way. The champion of scientific rationality, as you are rightly considered, has severely criticized the standard formulation of quantum theory and found it affected ... of subjectivism." (18/11/1984, PA, 348/21). Moreover Selleri, besides the abovementioned conferences, met Popper again in Naples in May 1985 (PA, 247/18) and even visited him at his home in February 1985 (Selleri to Popper, 31/01/1985). In 2003, almost ten years after Popper's passing, Selleri remembered: "In my opinion he was the best philosopher of the 20th century, the best philosopher of science, by far, without comparison. [...] I think that every time he was critical he was right".
    Another fact reported by some authors is that Popper and Selleri strictly limited their relation to the scientific level, due to their great political divergences. For instance, Freire writes: "Popper, celebrated critic of Marxism and advocate of liberalism, and Selleri, ostensible Marxists, opted for not including politics in the agenda of their conversations." (Freire, 2004, p. 118). They had indeed unequivocal different political positions, as it was also with Vigier who was as well a staunch communist (see Holland, 2004), but Selleri was surely interested in discussing these aspects as well, as showed by his assertion: "we only regret that there was not more time for [...] knowing your opinions on some problems of society, politics" (Letter to Popper on 16/05/1984, PA, 247/18).

39 Popper could not participate to this meeting, but he wrote a paper for the occasion, which was read by Bohm's former student, Chris Dewdney.

Meanwhile, because of the aftermath of Bari workshop, Popper's experiment started raising some attention in the community working on FQM.

In March 1984, Popper was in Vienna where he discussed his experiment with Helmut Rauch (who took part to the debate in Bari, and was the initiator of the interests towards FQM in Vienna in the post-war). Rauch then wrote a letter to Popper (02/05/1984, PA, 341/9) thanking him for the preprint of his essay on PE (the one to be published in the proceeding of the Bari conference) which he found very interesting. Moreover, on behalf of the *Chemisch-Physikalischen Gesellschaft* (Chemical-Physical Society), he invited Popper to offer a lecture on "Epistemology and Quantum Mechanics", in October.

It is also worth mentioning that Popper was again in Italy in May 1984, when he took part as a guest speaker in a conference organised in Naples (Italy) by the philosopher Gerardo Marotta to discuss the "new perspectives in quantum theory and in general relativity" (Baldini, 1998). There he had a new opportunity to deal with important physicists such as John Bell, Tullio Regge, Léon Van Hove, Robert Marshak and Bruno Zumino.

Other people started working on Popper's proposal, such as the philosopher of science (trained as a mathematical physicist) Henry Krips, who devoted most of his review of Popper's *Postscript to the Logic of Scientific Discovery* to a detailed analysis of PE (Krips, 1984, pp. 254-261). There he explicitly stated: "I predict (in opposition to Popper) that were it possible to perform the Popper experiment then we would find that $S_2(p_y) > S_1(p_y)$", meaning that the particle on the side where the slit had been removed should experience an increased scattering (see the possible outcomes predicted by Popper in section 4.3).

Another disagreement came from the mathematician Anthony Sudbery, who entered a critical paper, in July 1984, entitled "Popper's variant of the EPR experiment does not test the Copenhagen interpretation" ( Sudbery, 1985; see also Shields, 2012).

In April 1985, Selleri organised a conference in Cesena (Italy) to celebrate the 50th anniversary of the EPR experiment. In this case Popper did not give any contribution, but he was asked by Selleri to "help [...] in organising it" (19/09/1983, PA, 247/18), and to provide suggestions about the speakers to invite (letter from Selleri to Ivan Slade on 13/10/1984, PA, 348/21). At the conference, Gian Carlo Ghirardi - who was one of the most prominent Italian theoretical physicists in the field of FQM (see Baracca, Bergia and Del Santo, 2016, section 13.2.1) - presented a paper in which PE was criticised within the section "misunderstanding about the EPR analysis". There the author confirmed the notoriety that PE was acquiring, stating that "in view of the large interest that this book [Popper, 1982a] has created, we think it is useful to discuss its reasoning in detail." (Tarozzi and van der Merwe, 1988, pp. 95-98).

On the 25th of September 1985, a largely participated international conference (about two hundred physicists and philosophers of science), entitled *Microphysical reality and Quantum Formalism*, was organised in Urbino, Italy (Selleri and Tarozzi were again part of the organising committee). On that occasion Popper was officially a member of the international advisory committee and an invited speaker, although in the end he could not attend.[40] There, PE received certain consideration in a number of talks.

---

[40] Popper could not go to Urbino, because his wife was gravely ill (she passed away a few months later), but he arranged to send a contribution, published in the proceedings (Van der Merwe, Selleri And Tarozzi, 1988a, pp. 413-417). I am thankful to Prof. Gino Tarozzi for his testimony in a communication on 02/12/2016.

Sudbery enlarged his critiques to Popper's Experiment (van der Merwe, Selleri And Tarozzi, 1988a, pp.267-277). Whereas, the Belgian theoretical physicist Willy de Baere not only claimed that under certain conditions the experiment was possible in principle, but also that it could be used to "criticize state-vector reduction" (van der Merwe, Selleri And Tarozzi, 1988a, pp.425-429). Oreste Piccioni and Werner Mehlhop, from the University of California, also quoted PE as a remarkable example of simplicity, while complaining of the use of "complex formulas where few words [...] would suffice" (van der Merwe, Selleri And Tarozzi, 1988a, p. 383). Also Vigier, stimulated by the recent analysis of Bedford and Selleri, proposed a discussion on the "uncertainty relations (including $\Delta E \Delta t > \hbar$) at the origin of the particle pair separation", which he deemed it necessary for experiments of the kind of EPR, including that invented by Popper (van der Merwe, Selleri And Tarozzi, 1988b, p.211).

Popper also re-proposed his experiment in a selection of papers on foundations of physics in 1986 (Popper, 1986).

Meanwhile, the physicist Edward R. Pike drew the attention of his colleagues Matthew J. Collett and Rodney Loudon to PE, who published a letter on *Nature* in April 1987 in which they claimed that "the experiment cannot provide a crucial test of quantum mechanical interpretations" (Collett and Loudon, 1987). The authors, although they followed the tradition of basing their critiques upon the uncertainty at the source, also expressed their disagreement with the criticism to PE made by Sudbery (1985). Popper answered in a letter, published in *Nature* in the following August with the title "Popper versus Copenhagen" (Popper, 1987), providing additional arguments, which actually countered their objections. In particular, he stressed that Collett and Loudon would have agreed with his proposal provided a fixed source. *Nature* hosted another reply by Collet and Laudon (on the same issue) and a concluding note by Popper, but they did not reach an agreement (see Shields, 2012).

This marked the end of the debate on Popper's proposed experiment in the 1980s. The discussion on PE however burst again into physics journals only after 1999, when the experiment was finally conducted (Kim and Shih, 1999), though unfortunately five years after Popper had passed away.

*5.2 An overview on the experimental realisation of PE*

As it has been shown in the last section, the most severe critiques of PE were based on the impossibility of realising a suitable source of collinear quantum particles (e.g. photons). Even Selleri, albeit very much sympathetic with Popper's proposal, had proven that a two-body decay could not provide collinear entangled particles (see section 5.1).

It was only in the late 1980s that physicists started realising that a well known (at least since 1970) phenomenon of non-linear optics, the *Spontaneous Parametric Down Conversion* (SPDC), was producing entangled photons by its very nature (see e.g. Horne, Shimony and Zeilinger, 1990). This drastically simplified the way of producing entangled particles, and paved the way for a new era of experimental FQM, namely quantum optics.

One of the pioneers in this field is the physicist Yanhua Shih, who, together with Carrol Alley, was the first to exploit the SPDC in the realisation of experiments on quantum entanglement (Alley and Shih, 1986). Shih came to know of PE from Garuccio around 1995-1996 and immediately though he could have exploited the new techniques based on the SPDC (i.e. the so called *Ghost Imaging*) to solve the problem of the collinearity, completely avoiding the use of a point-like source (see Bromberg, 2001; Kim and Shih, 1999). So, Shih and his colleague Yoon-Ho Kim finally carried out, at the University of Maryland, the first experimental implementation of PE, and the results that they obtained were most astonishing: they found a complete agreement with Popper's theoretical predictions. The two experimentalists immediately drafted a paper entitled *Experimental Realization of Popper's Experiment: Violation of the Uncertainty Principle?* (Kim and Shih, 1999), which was published, not without difficulties (see section 5.2.1), in the journal

*Foundations of Physics*. Their answer to the question they had posed in the title of their paper was eventually negative, in fact Kim and Shih concluded that:

> Our experimental result is emphatically NOT a violation of the uncertainty principle which governs the behavior of an individual quantum.
>
> In both the Popper and EPR experiments the measurements are "joint detection" between two detectors applied to entangled states. Quantum mechanically, an entangled two-particle state only provides the precise knowledge of the correlations of the pair. Neither of the subsystems is determined by the state. (Kim and Shih, 1999, p.4)

Therefore in their opinion the mistake made by Popper is that of having applied the uncertainty principle separately for the two entangled particles and not to the couple together.

In any case, their outcome, although the conclusion seemed so preserve the CIQM, provoked the vigorous reaction of the community: more than a dozen papers on the matter appeared within the following three years (for the complete reference list see e.g. Shields, 2012). Moreover Kim and Shih had to defend their work from the many attacks within the physics community. The latter author recalled, as early as two years after their publication:

> I have been giving talks on Popper's experiment at least for three conferences, most of the people don't believe it. They don't believe me?!
>
> [...] the leaders in this field, was [*sic*] trying to argue with me. They said, "There must be something wrong there." (Bromberg, 2001)

Since then a number of research has been conducted trying to interpret the unexpected results of the experimental realisation of PE. It is beyond the scope of this paper, which covers specifically the decade of the 1980s, to reconstruct the controversy that followed the realisation of Popper's experiment. This has been already fully discussed both on a physical (e.g. Qureshi, 2012) and on a historical-epistemological perspective (e.g. Shields, 2012, section 5; Zouros, 2007).

However, the conclusion we can draw is that the vast majority of the most recent works (although not all of them) on PE, agrees on the fact that Heisenberg's uncertainty principle is indeed preserved, because it seems that also the Copenhagen Interpretation would lead, if properly calculated, to the same outcome (the first work firmly stating this is Qureshi, 2004). In this light, Popper's experiment basically seems not to be a crucial experiment to discriminate between a realist interpretation of quantum mechanics and Copenhagen's.

It is nevertheless interesting that 35 years after it was put forward, and 18 years after it was actually realised, PE still is matter of debate, having received new attention in recent years, either corroborating the thesis according to which it does not tests CIQM (Qureshi, 2012) or otherwise (Cardoso, 2015).

*5.2.1 A critical note on the modern debate*

A brief digression of methodological nature on the reaction of the physicists to the realisation of PE seems necessary and should deserve further attention.

In fact, when Kim and Shih submitted their paper containing the extremely interesting result of PE to two of the most influential journals, *Physical Review Letter* and *Nature*, they received a flat refusal from both of them. Shih himself remembered the episode in an interview:

> So we sent to Physical Review Letters, and it was rejected. Three referees. All of them said, "This is a violation of the uncertainty principle. It cannot be published." They said, "This is wrong. Basically it's wrong."
> [...] Yoon-Ho Kim also sent it to Nature. Nature also rejected it and said, "Well," basically the referees said they just don't like Popper's idea at all. They don't care what these formulas do; they just don't like it. Yeah. It was rejected. (Bromberg, 2001)

Indeed this marked the beginning of a battle conducted by the community of physicists against this unacceptable result. I am absolutely not questioning here the validity of the studies carried out on the subject, but surely none can benefit from the somewhat cheap way in which some authors dismissed Popper's proposal as trivial. As I have shown, the theoretical discussion around PE did not only puzzle the physicists for a decade, but the fact the PE represented a crucial experiment against CIQM had rarely been questioned on a theoretical basis, before its experimental implementation would show a disagreement with the orthodox expectations. And it should be appropriate to take this into account. Qureshi, in the most modern and comprehensive study on PE, has recently pointed out that

> none of the objections raised against Popper's experiment could convincingly demonstrate if there was a problem with the proposal. More surprisingly, Popper's inference that according to Copenhagen interpretation, localizing one particle should lead to the same kind of momentum spread in the other particle, was not refuted by anybody.
> (Qureshi, 2012, p. 2)

So, to easily dismiss PE as a non crucial test, even more after the experimental implementations have confirmed its most inconvenient outcome, and ignoring all the preceding theoretical investigations which agreed with the substantial validity of the test, seems in my opinion a scientific conduct to be avoided.

For instance Asher Peres, considered one of the founders of the modern quantum information, as early as in 1999 when the experiment was still underway, entered a paper in which, besides other questionable comments, he critically wondered why "Popper associates this absurd prediction (particle scatter due to potential knowledge by an observer) with the Copenhagen interpretation", without providing any concrete analysis (Peres, 1999).

Anthony J. Short also supported Copenhagen against Popper's attack, but with a different theoretical explanation (see. Shields, 2012, p. 8). A much more perplexing case is that of the recent paper "Popper's Thought Experiment Reinvestigated" (Richardson and Dowling, 2012), which opens with the sentence "Karl Popper posed an interesting thought experiment in 1934", wherein the authors inspect the consequences of the Kim and Shih's implementation of 'it' (obviously the performed experiment is definitely not the mistaken one, proposed by Popper in 1934, as the authors think). And if that weren't enough, they even conclude saying: "the fact that the results of Popper's thought experiment were never correctly calculated in the seventy plus years since it was proposed illustrates just how unintuitive and strange QM can be." Now, this is just a gross historical blunder, but it actually begs the question of which experiment has been 'reinvestigated' in this paper.

Besides these at best curious cases, one can claim that Popper's experiment was rather inconvenient and was therefore natural that it found strenuous opposition, since its very conception, of a vast the majority of physicists. But for a decade it was confuted mostly on the basis of its mere feasibility, not questioning its being crucial. Yet, as soon as PE became feasible and it had been conducted, suddenly its result turned out to be unacceptable to most of physicists. This instilled new motivation to look for a fatal flaw in Popper's proposal, on a fundamental basis. Since many different explanations, often in contradiction between each other, had appeared soon after PE was conducted, and more are today available, this might suggest that research of such a kind can be biased by prejudices (at least unconsciously). So, before jumping into the research of a 'necessary' criticism, a methodological consideration seems essential, and a serious historical account on the preceding works on the subject cannot hurt.

Selleri, concerning the analogous EPR paradox, stated that "one has to be careful because this matter is strongly steeped in philosophy, thus in passions, and some distortions have been introduced by various authors." (Selleri, 1989, p.185)

*5.3 Further Popper's contributions to QM.*

Besides PE, which still represents the most concrete and important contribution of Popper to QM, I deem it necessary to at least mention the other activities concerning FQM, that Popper carried out throughout the 1980s.

It is worth mentioning that Popper was aware of Bell's inequalities as early as in 1969 (letter to Abner Shimony 30/11/1969 PA, 350/7), at a time where that result did not cause almost any response in the physics community. But it is during the eighties that he started criticising the 'universality claim' of Bell's theorem, basing his criticisms on the alleged confutation of a demonstration provided by Clauser and Horne (1974). Popper thus published two works: "Towards a Local Explanatory Theory of the Einstein-Podolsky-Rosen-Bohm Experiment" (Popper and Angelidis, 1985), and "Bell's theorem: a note on locality" (Van der Merwe, Selleri And Tarozzi, 1988a, pp. 413-417). Indeed, Popper never abandoned the idea that "there is no reason to regard Q.M. as non-local. It is a local theory. It is, of course, incomplete." (letter to Selleri, 30/11/1983, PA, 348/21). As to this line of research, Selleri concluded that Popper, "in his eighties, started again doing the theoretical physicist with youthful enthusiasm trying to destroy Bell's theorem, but without concrete results" (Selleri, 1989, p. 186).

A further pivotal contribution of Popper to the foundations of quantum physics, is the formulation of an alternative statistical interpretation of quantum formalism. Indeed, Popper, who was a supporter of indeterminism (see section 4.2), created a sort of objective probabilistic realism, based on the idea of what he called *propensities* (Popper, 1957; Popper, 1989). This interpretation won the support of some physicists, like Leslie Ballentine, who had worked on statistical interpretation of QM (Ballentine, 1970), and was fully embraced by Selleri, who about this wrote to Popper:

> Your propensity interpretation of probabilities and your reinterpretation of Heisenberg's relations are for me natural. (07/12/1983, PA, 348/21)

The resonance of Popper's propensity interpretation among physicists had been analysed in detail by Olival Freire (2004), with an emphasis on the figure of Selleri. Also the correspondence between Selleri and Popper (PA, 562/19) and some of Selleri's works (e.g. Selleri, 1989, pp. 180-182) deal with this subject.

As a further acknowledgment by the scientific community, Popper was also appointed member of the advisory committees of several international conferences on FQM, together with the most illustrious physicists in the field. This happened, for instance, for the abovementioned conferences in Cesena (1985), in Urbino (1985), and for a conference in Delhi, organised in September 1988 by Ranjit Nair (Letter from Selleri on 26/07/1988, PA, 562/19). Moreover Popper acquired in 1988 the membership into the scientific committee of the then recently created journal *Foundation of physics Letters*, edited by Alwyn van der Merwe (PA, 562/19).

Popper also had a second scientific collaboration with Vigier, which led to a new co-authored paper together with the Serbian physicist Zvonko Marić (Marić, Popper and Vigier, 1988).

As the decade we have considered drew to a close, Popper's remarkable involvement in QM drastically decreased. Besides his old age (he was almost ninety), this probably happened also because of the dissatisfaction of not having seen his own experiment realised.[41] He wrote a last letter to Selleri on the 11th of December 1989, of which the conclusive words give us a taste of Popper's tireless devotion

> I am getting extremely old but I still have new and exciting ideas in various fields [...], and I am happy. (PA 562/19)

**ACKNOWLEDGEMENTS**

I am indebted to the *Karl Popper Sammlung* of the AAU in Klagenfurt for granting me the access to their archives and the reproduction of Popper's original correspondence and drawings, in particular to Mag.[a] Nicole Sager and Dr. Manfred Lube for their irreplaceable help.

I am warmly grateful to Mr. David Miller for his continuous support and the many valuable suggestions and to Prof. Karl Milford for his kind encouragement.

I would like to thank Prof. Olival Freire Jr. for the kind comments that greatly improved this paper.

I am also thankful to Prof. Joseph Agassi, Prof. Nicola Cufaro Petroni, Prof. Basil Hiley and Prof. Gino Tarozzi for the interesting correspondence.

I would like moreover to thank Ms. Azzura Sorbi and Mr. Daniel Long for the precious revision of the English language of the manuscript.

This research did not receive any specific grant from funding agencies in the public, commercial, or not-for profit sectors.

---

41 Popper was interviewed a last time in 1992 about the EPR controversy and still he affirmed that his experiment "should be done" (Combourieu, 1992, p. 9). There Popper also mentioned that he had received some proposals from experimentalists (fom Canada and New Zeeland) who approached him to possibly perform PE, but he complained about their scientific attitude, concluding: "I cannot go ahead with people who want to get money". It is moreover worth mentioning that Popper did not devote further attention to FQM, also because in his late years he became more and more involved with Wächtershäuser's work on the origin of life (see e.g. Miller, 1997). I am grateful to Mr. David Miller for having pointed out to me this change in Popper's interests (personal communication on 11/01/2017).